\documentclass[aps,prd,reprint,preprintnumbers,superscriptaddress,showpacs,twocolumn]{revtex4-1}
\usepackage{latexsym,graphicx,amssymb,amsmath,mathrsfs}
\usepackage{setspace,bm}
\usepackage[breaklinks, colorlinks=true, pdfstartview=FitV, linkcolor=red, citecolor=blue, urlcolor=blue]{hyperref}
\usepackage[usenames]{color}
\usepackage{latexsym}
\usepackage{epstopdf}
\usepackage{mathtools}
\usepackage{url}
\usepackage{comment}
\usepackage{braket}
\usepackage{CJKutf8}

\begin{document}

\title{Application of the path optimization method to a discrete spin system}

\author{Kouji Kashiwa}
\email[]{kashiwa@fit.ac.jp}
\affiliation{Department of Computer Science and Engineering, Faculty of Information Engineering, Fukuoka Institute of Technology, Fukuoka 811-0295, Japan}

\author{Yusuke Namekawa}
\affiliation{Education and Research Center for Artificial Intelligence and Data Innovation, Hiroshima University, Hiroshima 730-0053, Japan}

\author{Akira Ohnishi}
\affiliation{Yukawa Institute for Theoretical Physics, Kyoto University, Kyoto 606-8502, Japan}
\email{Akira Ohnishi passed away on May 16th 2023, while this paper was in preparation.}

\author{Hayato Takase}
\noaffiliation{}

\begin{abstract}
The path optimization method, which is proposed to control the sign problem in quantum field theories with continuous degrees of freedom by machine learning, is applied to a spin model with discrete degrees of freedom.
The path optimization method is applied by replacing the spins with dynamical variables via the Hubbard-Stratonovich transformation, and the sum with the integral.
The one-dimensional (Lenz-)Ising model with a complex coupling constant is used as a laboratory for the sign problem in the spin model.
The average phase factor is enhanced by the path optimization method, indicating that the method can weaken the sign problem.
Our result reproduces the analytic values with controlled statistical errors.
\end{abstract}
\maketitle

\section{Introduction}

To understand the non-perturbative properties of quantum field theories and spin models, the Monte Carlo (MC) method plays an important and crucial role.
In the MC calculation, expectation values are evaluated with the Boltzmann weight.
However, the Boltzmann weight is a complex value in some cases even if the partition function is still real.
This problem is called the sign problem.
A typical example is quantum chromodynamics with a finite quark chemical potential ($\mu$), reviewed in Refs.\,\cite{deForcrand:2010ys,Nagata:2021bru,*Nagata:2021ugx}.
Another example of a discrete spin system is the Hubbard model away from the half-filling~\cite{loh1990sign}.

The sign problem can be milder by the path optimization method or the sign-optimized manifold~\cite{Mori:2017nwj,Alexandru:2018fqp,Bursa:2018ykf}, which has a close relationship with the Lefschetz thimble method~\cite{Witten:2010cx}.
Both are the so-called complexified dynamical variable approaches based on Cauchy's integral theorem, which ensures the independence of the expectation value via modification of the integral path as long as the integrand is an entire function with no contribution at infinity.
If we have an integral representation of the partition function, such as quantum field theories with continuous degrees of freedom, a sign-problem-reduced integral path could be on the complexified dynamical variable plane.
The path optimization method utilizes machine learning to determine the optimized integration path.
The path optimization works well for several models: a simple Gaussian model~\cite{Mori:2017nwj}, the $1+1$ dimensional complex $\lambda \phi^4$ theory~\cite{Mori:2017pne}, the Polyakov-loop extended Nambu--Jona-Lasinio model~\cite{Kashiwa:2018vxr,Kashiwa:2019lkv}, the 1+1 and 2+1 dimensional Thirring model~\cite{Alexandru:2018fqp,Alexandru:2018ddf}, the $0+1$ dimensional Bose gas~\cite{Bursa:2018ykf}, the $0+1$ dimensional QCD~\cite{Mori:2019tux}, the two-dimensional $\text{U}(1)$ gauge theory with complexified coupling constant~\cite{Kashiwa:2020brj,Namekawa:2021nzu,Namekawa:2022liz}, the 2+1 dimensional XY model~\cite{Giordano:2022miv}.
It is also employed for error reduction of observables~\cite{Detmold:2020ncp,Detmold:2021ulb}.
The recent progress of the complexified dynamical variable approach is reviewed in Ref.~\cite{Alexandru:2020wrj}.

For the spin models, on the other hand, we have a sum in the partition function, instead of the integral.
We cannot directly apply the complexified dynamical variable approach.
A solution is the Hubbard-Stratonovich transformation.
It converts the sum to the integral using the auxiliary field.
We demonstrate it in the one-dimensional classical (Lenz-)Ising model with a complex coupling constant.
Since we have the analytic result, we can judge the correctness of results by the path optimization method.
Another famous example is the Hubbard model away from half-filling~\cite{loh1990sign}.
In this case, the complexified dynamical variable approach is feasible; for example, see the review~\cite{Berger:2019odf}.

In this paper, we apply the path optimization method~\cite{Mori:2017pne,Mori:2017nwj} to the Ising model with the complex coupling constant.
The dynamical variables are replaced by the Hubbard-Stratonovich transformation as in Refs.\,\cite{NIPS2012_c913303f,Ostmeyer:2019dkt}.
We also introduce parallel tempering to the path optimization method~\cite{Kashiwa:2020brj} toward control of the global sign problem, as first applied in the Lefschetz thimble method~\cite{Fukuma:2017fjq,Fukuma:2019wbv}.

This paper is organized as follows.
In Sec.~\ref{sec:formulation}, we explain the formulation of the Ising model with the complex coupling constant, the Hubbard-Stratonovich transformation, and the path optimization method.
The numerical setup and results are shown in Sec.~\ref{sec:setup} and Sec.~\ref{sec:results}, respectively.
Section~\ref{sec:summary} is devoted to a summary.

\section{Formulation}
\label{sec:formulation}

We employ the one-dimensional classical Ising model with a complex coupling constant as a laboratory to investigate the sign problem in spin models.
The sign problem is induced by the imaginary part of the external field.
We first explain the integral representation of the Ising model through the Hubbard-Stratonovich transformation.
We then explain the application of the path optimization method to the model.

\subsection{Ising model with complex coupling constant}

The Hamiltonian of the classical one-dimensional (Lenz-)Ising model with an external magnetic field~\cite{lenz1920beitrag,Ising:1925} is given by
\begin{align}
    {\cal H} &= -J \sum_i \sigma_i \sigma _{i+1} - h \sum_i \sigma_i,
\end{align}
where $J$ is a coupling constant for the nearest-neighbor spins, $h$ is strength of the external magnetic field, and $\sigma_i = \pm 1$ is a spin at each site $i=1,\cdots,N$; this is the one-dimensional Ising chain.
We impose the periodic boundary condition, $\sigma_{0}=\sigma_{N}$ and $\sigma_{N+1}=\sigma_1$.
The Hamiltonian ${\cal H}$ can be represented in matrix style as
\begin{align}
    {\cal H} & = - \frac{J}{2} s^\top K s - H s,
    \label{eq:Ising2}
\end{align}
where $K$ is the symmetric connectivity matrix, $s$ is the spin matrix defined as $s=(\sigma_1~\sigma_2~\cdots~\sigma_N)^\top$, and $H = h \times (1~1~\cdots~1)$.
This representation can be also applied to higher-dimensional systems by using a suitably constructed symmetric connectivity matrix.
The coefficient $1/2$ in Eq.\,(\ref{eq:Ising2}) is introduced to avoid double counting of the nearest-neighbor interaction when we make $K$ symmetric.

The sign problem arises from the imaginary part of $J$.
The realistic Ising model does not have such an imaginary part, but is sometimes introduced for analysis of the Lee-Yang zeros~\cite{lee1952statistical} or the Fisher zeros~\cite{fisher1965statistical}.
Such an imaginary part also naturally arises when we consider the QCD-like Potts model~\cite{Alford:2001ug,Kim:2005ck,Kashiwa:2020waa}, as discussed in Appendix\,\ref{sec:QCD_like_Potts}.

The partition function of the Ising model is
\begin{align}
    {\cal Z} = \sum_{\{s_i\}=\pm 1} e^{- \beta {\cal H}(J, h)} = \sum_{\{s_i\}=\pm 1} e^{-{\cal H}(J', h')},
\end{align}
where the sum takes over all possible states.
The inverse temperature $\beta$ can be absorbed into ${\cal H}$ by replacing $J'=\beta J$ and $h'=\beta h$.

\subsection{Hubbard-Stratonovich transformation}

With the expression (\ref{eq:Ising2}), we can use the Hubbard-Stratonovich transformation as in Ref.\,\cite{Ostmeyer:2019dkt};
\begin{align}
    e^{\frac{1}{2}s^\top K s}
    = \frac{1}{\cal N}
      \int_{-\infty}^{\infty} \Bigl[ \prod_{j=1}^{N} d v_i \Bigr]
      e^{-\frac{1}{2} v^\top K^{-1} v
      + s \cdot v},
\end{align}
where the normalization constant ${\cal N}$ is defined by
\begin{align}
    {\cal N} = \sqrt{(2\pi)^N \det K}.
\end{align}
It should be noted that the eigenvalue of $K$ must be positive for the Hubbard-Stratonovich transformation.
We thus put a constant shift for $K$ as
\begin{align}
    K \to {\tilde{K}} = K + C {\bf I},
 \label{eq:shift_K}
\end{align}
where ${\bf I}$ is the unit matrix and the constant $C$ takes the same sign as that of $J$.
The $C$-independence of the physical result is confirmed in Ref.\,\cite{Ostmeyer:2019dkt}.
If we set $C > n$ where $n$ is the maximum number of nearest neighbors of one site, ${\tilde K}$ is positive definite;
$n=2$ for the one-dimensional Ising model.

The final form of the partition function becomes
\begin{align}
    {\cal Z}
    &= \sum_{\{\sigma_i=\pm 1\}} e^{-{\cal H} - \frac{J'}{2}C s^2}
    \nonumber\\
    &= \frac{1}{\cal N'} \int_{-\infty}^\infty
    \Bigl[ \prod_{j=1}^{N} d v_i \Bigl]
    e^{ -\frac{1}{2J'} v^\top {\tilde{K}}v
    + \sum_j \ln \cosh[H'_j+({\tilde K}v)_j]} 
    \nonumber\\
    &= \frac{1}{\cal N'} \int_{-\infty}^\infty
    \Bigl[ \prod_{j=1}^{N} d v_i \Bigl]
    e^{ -{\cal H}'},
\end{align}
where ${\cal N'}$ includes a contribution of ${\cal N}$ and $C$, which is irrelevant in the evaluation of the expectation values.
We can consider ${\cal H}'$ as the effective Hamiltonian in molecular dynamics.
The expectation value of magnetization for the single spin is obtained as
\begin{align}
    \langle {\sigma} \rangle &= \Bigl\langle \frac{1}{N} \sum_j \tanh \Bigl[ H'_j+({\tilde K}v)_j \Bigr] \Bigr\rangle.
    \label{eq:magnetization}
\end{align}

The analytic result of the magnetization~\cite{Ising:1925} is known as
\begin{align}
    \langle \sigma \rangle &= \frac{\lambda_+^N - \lambda_-^N}{\lambda_+^N + \lambda_-^N} \frac{\sinh(h')}{\sqrt{\sinh^2(h') + e^{-4J'}}},
\end{align}
where $\lambda_\pm$ are the eigenvalues of the transfer matrix of the model,
\begin{align}
\lambda_\pm = e^{J'} \Bigl[ \cosh(h') \pm \sqrt{\sinh^2(h')+e^{-4J'}} \Bigr].
\end{align}

\subsection{Path optimization method}

The path optimization method~\cite{Mori:2017nwj,Alexandru:2018fqp,Bursa:2018ykf} is proposed as a complex dynamical variable approach for the path integral formulation to control the sign problem via machine learning.
Although the path optimization method does not need initial teacher data, the effectiveness of the modified path can be automatically evaluated in the learning part.

In the path optimization method, we first complexify the dynamical variable $v \in \mathbb{R}^N$ as
\begin{align}
    v \to v' = v_\mathrm{R} + i v_\mathrm{I},
    \label{eq:modified_path}
\end{align}
where $v_\mathrm{R}, v_\mathrm{I} \in \mathbb{R}^N$.
This procedure means modification of the integral path on the complexified dynamical variable plane.
There are several ways to express the modified integral path minimizing the sign problem.
We use the representation constructed by the neural network;
the input is $v = v_R$, and the output is $v_I$.
The actual procedure is as follows:
\begin{align}
    \underbrace{v}_{\mathrm{input\,layer}} \to \mathrm{hidden~layer}
      \to \underbrace{v_\mathrm{I}}_{\mathrm{output\,layer}}.
\label{eq:layer}
\end{align}
The output layer is
\begin{align}
    v_{\mathrm{I}, l} = v^{(L)}_l &= [ w_{lk}^{(L-1)} f(v_k^{L-1}) + b_l^{(L-1)}],
\end{align}
where $L$ is the total number of layers.
The hidden layer is composed of
\begin{align}
    v_k^{(l+1)} &= w_{kj}^{(l)} v_j^{(l)} + b_k^{(l)},
\end{align}
where
\begin{align}
    v^{(l)}_j &= w_{ji}^{(l-1)} f(v_i^{(l-1)}) + b_j^{(l-1)}.
\end{align}
$v^{(l)}$ indicates quantities on the $l$-th layer ($l=1,\cdots,L-1$) with $v^{(0)}=v$.
Weight $w$ and bias $b$ are the parameters of the neural network optimized by the back-propagation method with the appropriate cost function.
The activation function is the hyperbolic tangent, $f(\cdot) = \tanh(\cdot)$.
In this work, we employ the following cost function with and without a penalty term explained in Sec.~\ref{sec:extensions}:
\begin{align}
    {\cal F}(w,b) &= \int d v_\mathrm{R} \, |e^{i\theta(v_\mathrm{R})}-e^{i \theta_0}|^2 \,
    |{\cal J}(v_\mathrm{R}) \, e^{-S(v')}|, 
    \label{eq:cost_function}
 \\
 e^{i \theta(v_\mathrm{R})} &= {\cal J}(v_\mathrm{R}) e^{-S(v')} / |{\cal J}(v_\mathrm{R}) e^{-S(v')}|,
\end{align}
where ${\cal J}(v_\mathrm{R})$ is Jacobian, $S(v')$ represents the action composed of $v'$.
$\theta_0$ means the phase of the partition function.
Since we do not know the exact value of $\theta_0$, we estimate it iteratively in the learning process. 

The actual procedure is as follows:

\begin{enumerate}
 \item Generate configurations on the original path
 \item Update the neural network parameters using the generated configurations
 \item Regenerate configurations on the modified path
 \item Repeat 2 and 3 to obtain a converged result
\end{enumerate}

Since the Boltzmann weight is still complex even on the modified path, phase reweighting is required for the probability:
\begin{align}
    \langle {\cal O} \rangle
    &= \frac{\langle {\cal O} e^{i\theta} \rangle_\mathrm{pq}}
            {\langle e^{i\theta} \rangle_\mathrm{pq}},
\label{eq:pq}
\\
    \langle {\cal O} \rangle_\mathrm{pq}
&:= \frac{1}{{\cal Z}_\mathrm{pq}} \int  dv_\mathrm{R} \,  {\cal O} \, |{\cal J}(v_\mathrm{R}) e^{-S(v')}|,
\end{align}
where ${\cal O}$ represents an observable such as magnetization.
The left-hand side in Eq.\,(\ref{eq:pq}) is the correct expectation value of ${\cal O}$, while $\langle \cdots \rangle_\mathrm{pq}$ is the phase-quenched expectation value, and ${\cal Z}_\mathrm{pq}$ is the partition function with the corresponding Boltzmann weight.
The denominator of Eq.\,(\ref{eq:pq}) is the so-called average phase factor (APF).
If APF is exactly $1$, the sign problem completely disappears.
The sign problem becomes serious when the APF approaches $0$.

Note that the path optimization method and other sign-optimized manifold approaches usually require a Jacobian calculation to modify the integral path, which requires a high numerical cost, ${\cal O}(N^3)$.
In this work, we consider the simple model, and thus do not introduce the reduction technique of the Jacobian calculation, but we need it for more complicated models and theories.
One of the possible ways is that we completely neglect the Jacobian calculation in the learning part; this is a most drastic reduction technique because the Jacobian is completely neglected except in the evaluation part of the expectation values.
In Ref.~\cite{Namekawa:2022liz}, such a drastic approximation is shown to work at least in the $1+1$ dimensional $U(1)$ gauge theory.
Another treatment of the reduction of the Jacobian calculation, for example, is discussed by using the affine coupling layer in Ref.\,\cite{Rodekamp:2022xpf}.
No Jacobian calculation is required in the configuration generation of the worldvolume Hybrid Monte Carlo method by use of the flow equations~\cite{Fukuma:2020fez,Fukuma:2021aoo}.

\subsection{Parallel tempering}
Since the path optimization method makes the phases of the Boltzmann weight in the partition function localize, we may encounter the global sign problem even if it seems to be absent on the original integral path.
The global sign problem arises if there are some relevant contributions on the integral path which are separated by the energy barrier in the molecular dynamics.
To treat the global sign problem, we consider the parallel tempering method \cite{swendsen1986replica,geyer1991markov,hukushima1996exchange}, as adopted to the Lefshetz thimble method \cite{Fukuma:2017fjq} and the path optimization method \cite{Kashiwa:2020brj}.

In this study, we introduce replicas as follows, instead of varying the temperature.
We modify the integral path using the path optimization method, and we have Eq.\,(\ref{eq:modified_path}).
We then make replicas as
\begin{align}
    v'_r = v_\mathrm{R} + i \frac{v_\mathrm{I}}{r},
    \label{eq:replicas}
\end{align}
where $r = 1, \cdots, N_r$ and $N_r$ means the total number of replicas.
The region between the original path and the modified path is divided into $N_r$ slices as replicas. 
The exchange probability between the $r$th-replica and $(r+1)$th-replica is set as
\begin{align}
  {\cal P} 
  &= \min \left( 1,
     \frac{ {\cal P}(v'_{r}; r+1) {\cal P}(v'_{r+1}; r)}
          { {\cal P}(v'_{r}; r) {\cal P}(v'_{r+1}; r+1)}
          \right),
\end{align}
where
\begin{align}
  {\cal P} (v'_{r}; r)
  &= {\cal J}(v'_{r}; r) \, e^{-\mathrm{Re}\,H' (v'_{r};r)}.
\end{align}

\subsection{Improvements}
\label{sec:extensions}

We introduce the following three improvements to the path optimization method.
Improvements are in part based on knowledge obtained in the machine learning community. 

First, we add the penalty term to the cost function~(\ref{eq:cost_function}), similar to the $L_2$ normalization,
\begin{align}
 {\cal F}_\mathrm{penalty}
    &= \lambda \int dv_\mathrm{R} \,
    \sum_{i=1}^N \frac{\Bigl( \mathrm{Im}\,v_i(v_\mathrm{R}) \Bigr)^2}{N},
    \label{eq:cost_function2}
\end{align}
where $\lambda$ is the strength of the term.
The penalty term prohibits too large separations of the integral path from the original path in the training.

Second, we mix the previous and regenerated configurations as $50:50$ in the training part to make the training speed moderate; we only use the regenerated configurations in the evaluation of the expectation value.
If the regenerated configurations are significantly changed compared with the previous configurations, it may violate the stability of training.

Finally, we introduce the scheduler, ExponentialLR~\cite{li2019exponential}, to ensure stable training.
The scheduler decreases the learning rate as the training progresses, and thus the change of the parameters in the neural network becomes mild.
If the model approaches good minima, a decrease in the learning rate leads to better determination of the parameters.

\section{Numerical setup}
\label{sec:setup}

\begin{figure}[t]
 \centering
 \includegraphics[width=0.235\textwidth]{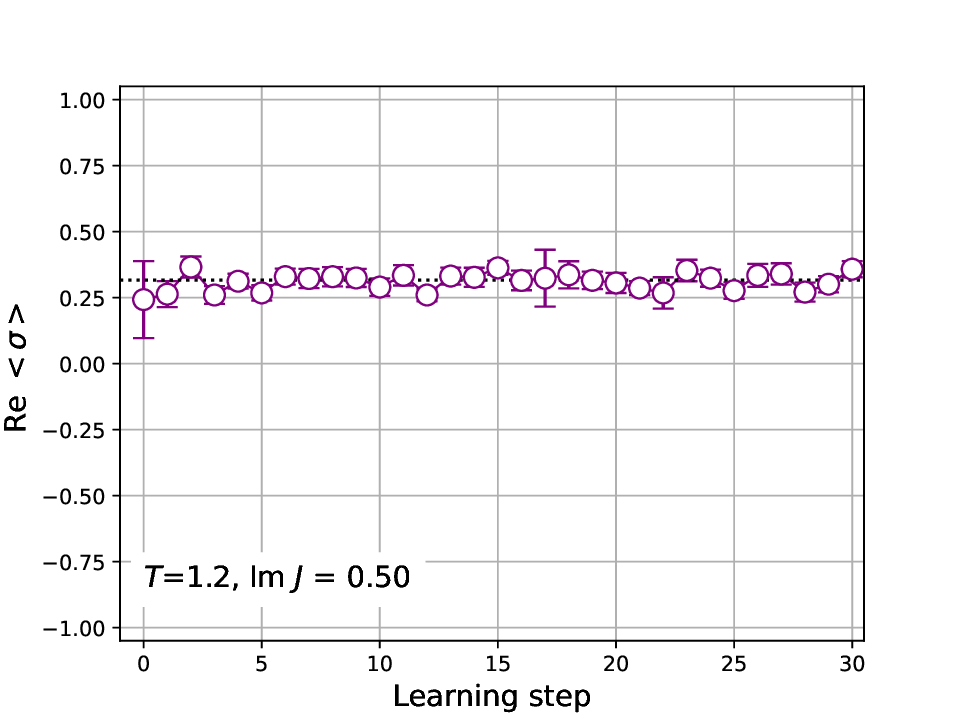}
 \includegraphics[width=0.235\textwidth]{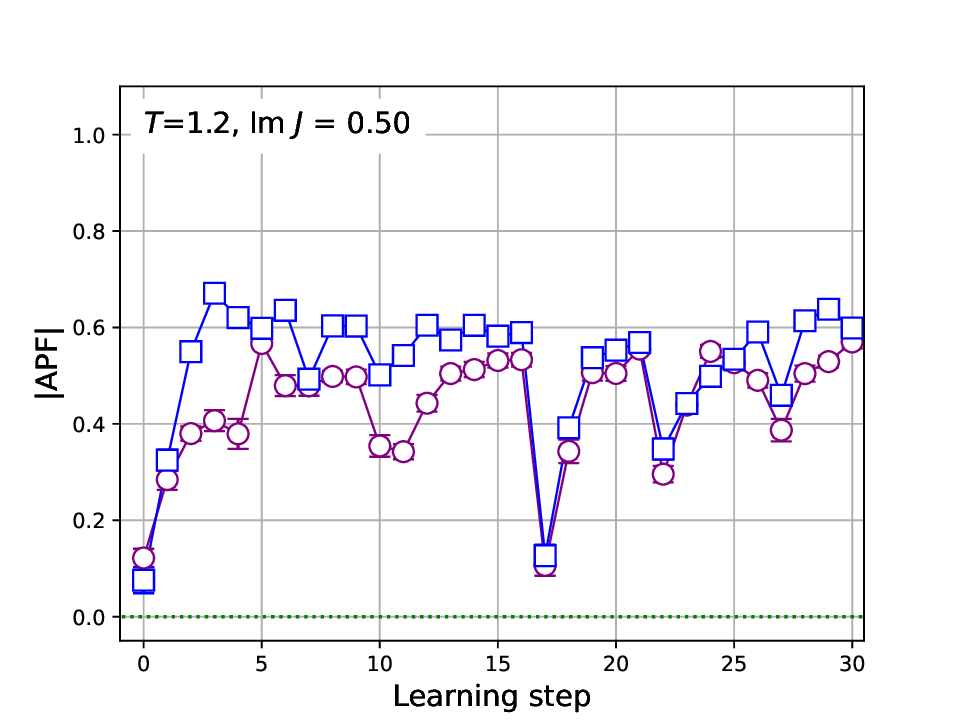}\\
 \includegraphics[width=0.235\textwidth]{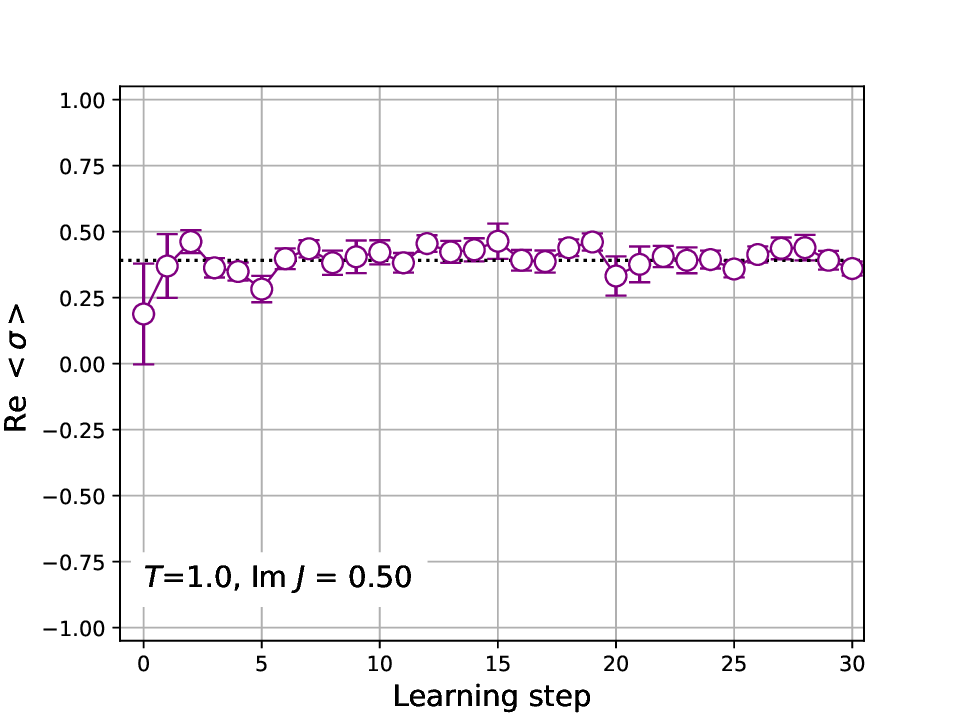}
 \includegraphics[width=0.235\textwidth]{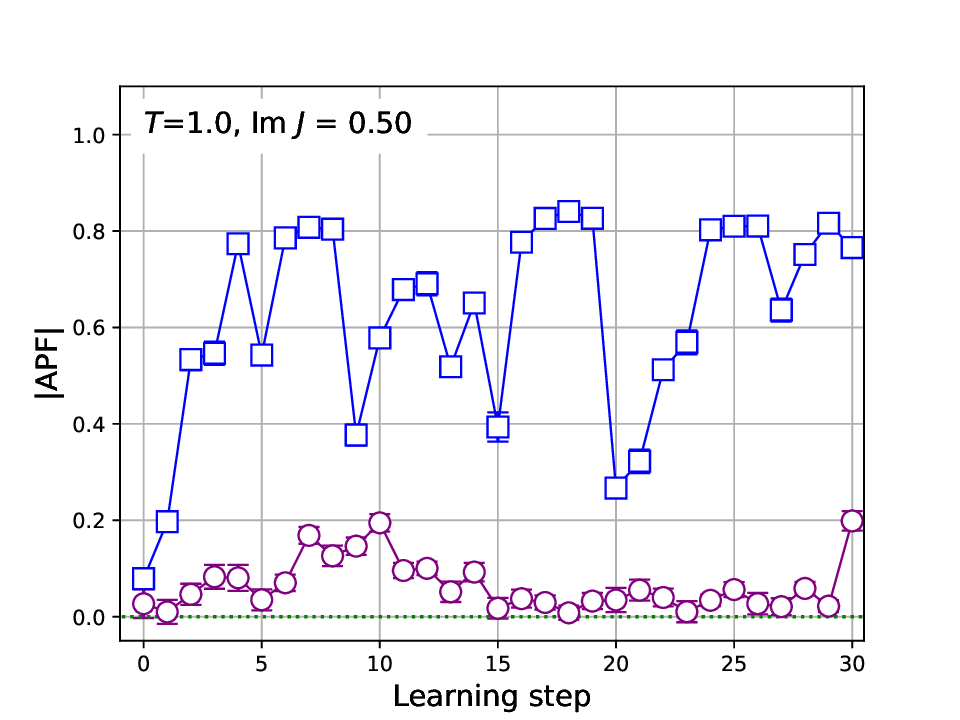}\\
 \includegraphics[width=0.235\textwidth]{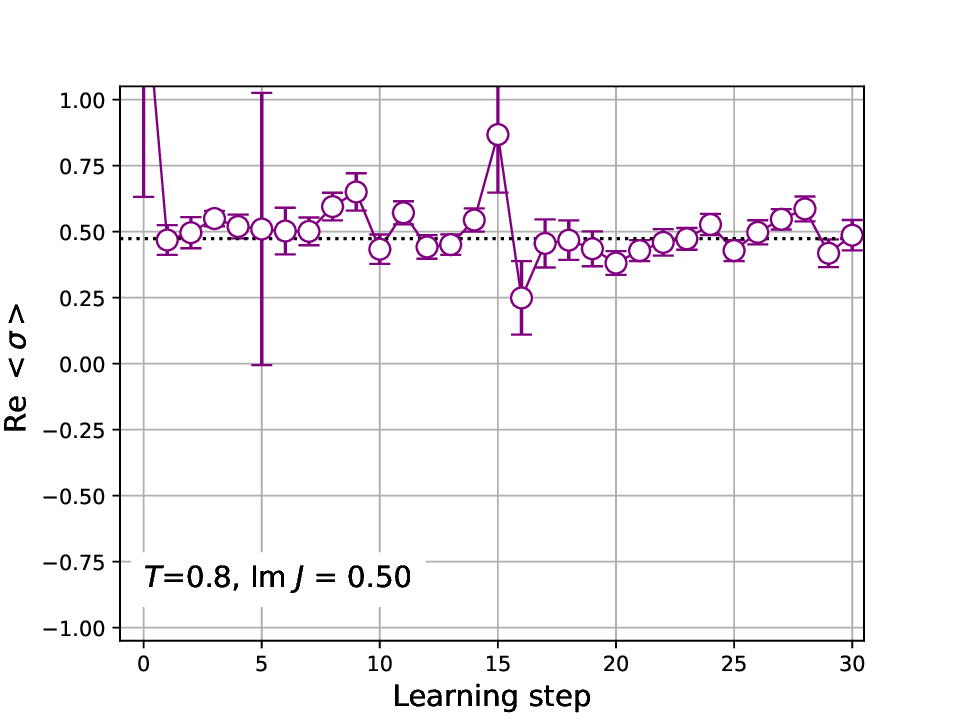}
 \includegraphics[width=0.235\textwidth]{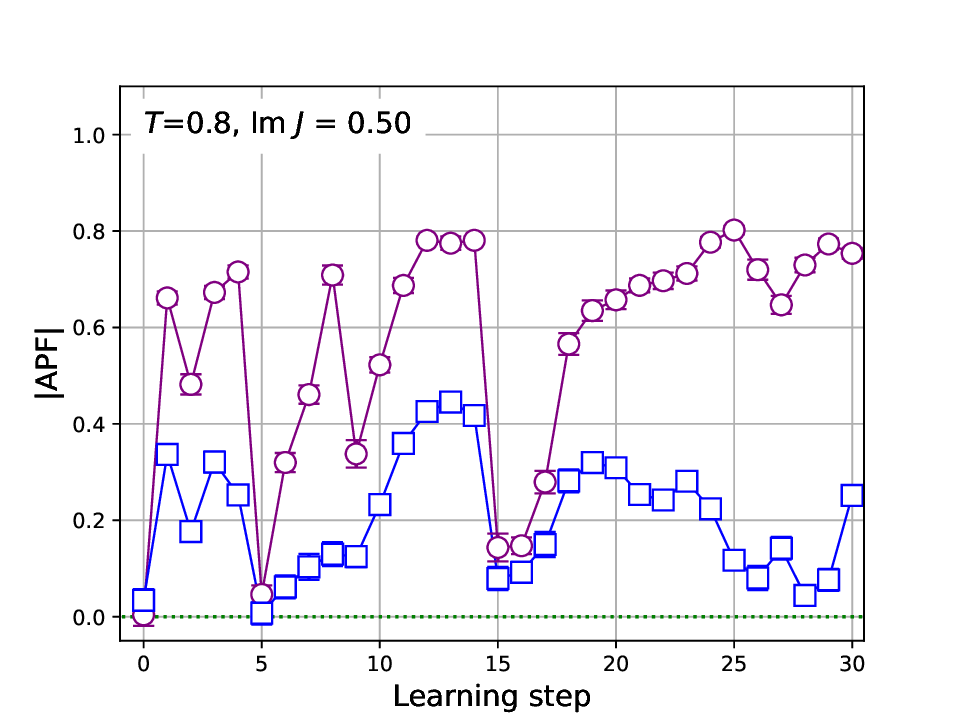}\\
 \caption{The magnetization and APF with $\mathrm{Im}\,J = 0.5$ at $T=0.8$, $1.0$ and $1.2$.
          Here, we do not introduce three improvements.
          The left panel shows the real part of the magnetization and the right panel shows APF.
          The circle and square symbols in the right panel are the results of the real and imaginary parts of APF, respectively.}
\label{fig:ImJ-simple}
\end{figure}

We consider $N=4$ spins in the one-dimensional Ising model.
Our numerical codes are implemented in the framework of PyTorch~\cite{paszke2019pytorch}.
For evaluation of the expectation values, we generate $N_\mathrm{conf}=1000$ configurations after thermalization by HMC.
The trajectory length is 1 with a step size of 0.2.
For the number of replicas, we employ $N_\mathrm{r} = 10$.
The statistical error is estimated using the Jackknife method with bin size $50$.
Measurements are performed at each $100$ trajectories.
We set $\mathrm{Re}\,J=1.0$ and $h=0.1 \in \mathbb{R}$.
The shift value $C$ in ${\tilde K}$ in \eqref{eq:shift_K} is $2+10^{-5}$.

In the training part, we use the batch training~\cite{bottou1998online} with the batch size $32$.
The number of hidden layers is $L=2$ and each layer contains $64$ units.
We employ AdamW~\cite{loshchilov2017decoupled} as an optimizer.
We set the strength of the penalty term $\lambda=1.0$.
The decay rate on the scheduler is $\gamma = 0.9$.
In the following, we show the results with the three improvements explained in Sec.\,\ref{sec:extensions}.
The results are evaluated after the $30$th training.
If no significant improvement is achieved in the early stage of training, we reset the initial values of the neural network.

After finishing the training, we regenerate the configurations and estimate APF and the magnetization.

\section{Numerical results}
\label{sec:results}

\begin{figure}[t]
 \centering
 \includegraphics[width=0.235\textwidth]{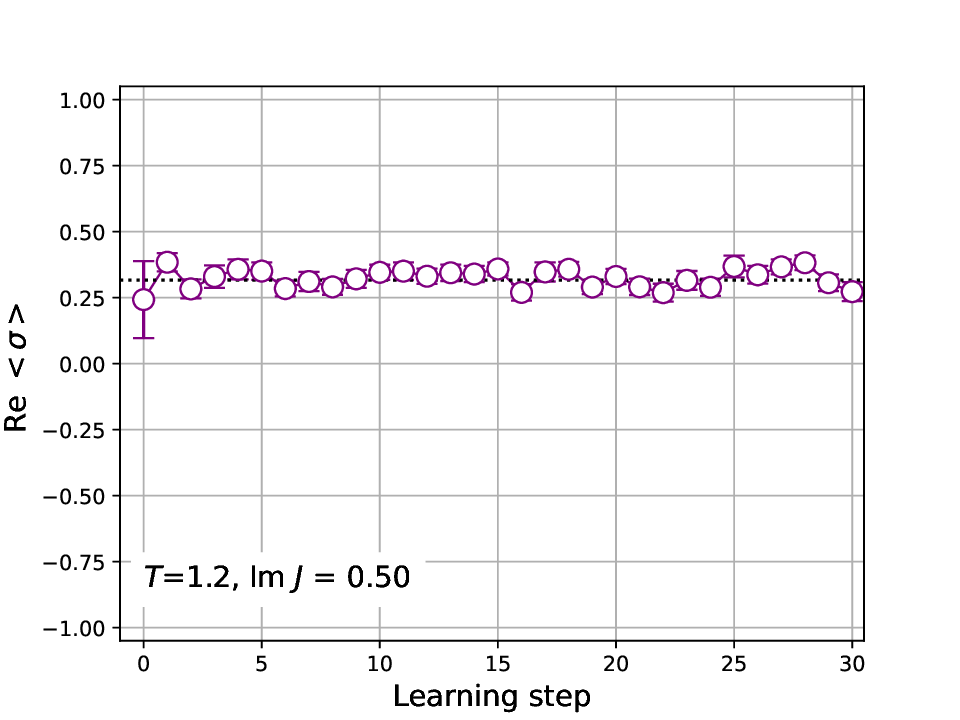}
 \includegraphics[width=0.235\textwidth]{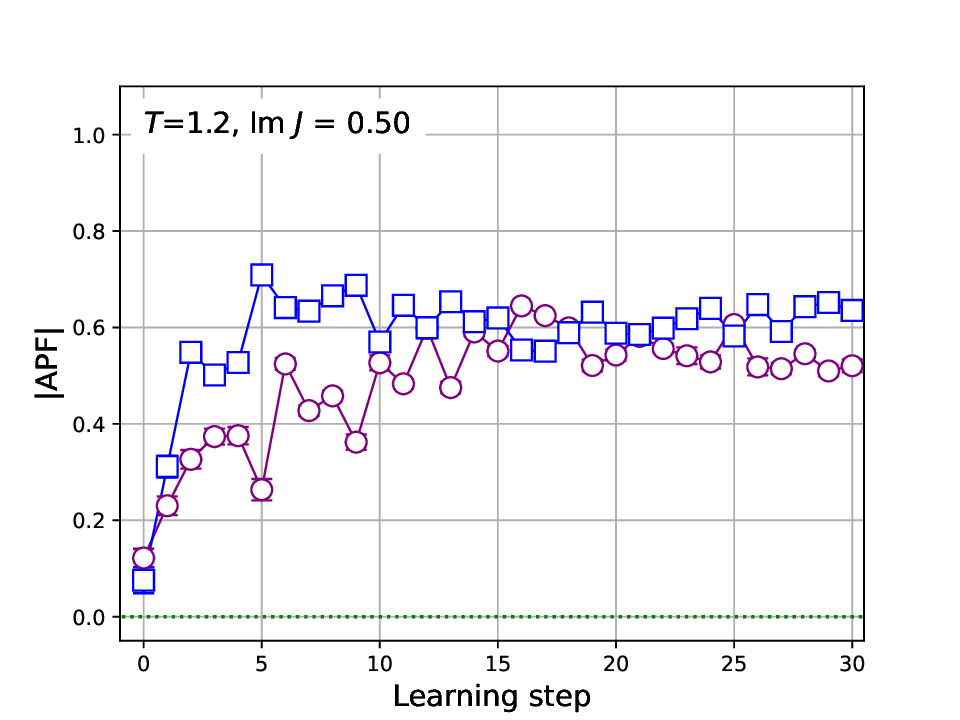}\\
 \includegraphics[width=0.235\textwidth]{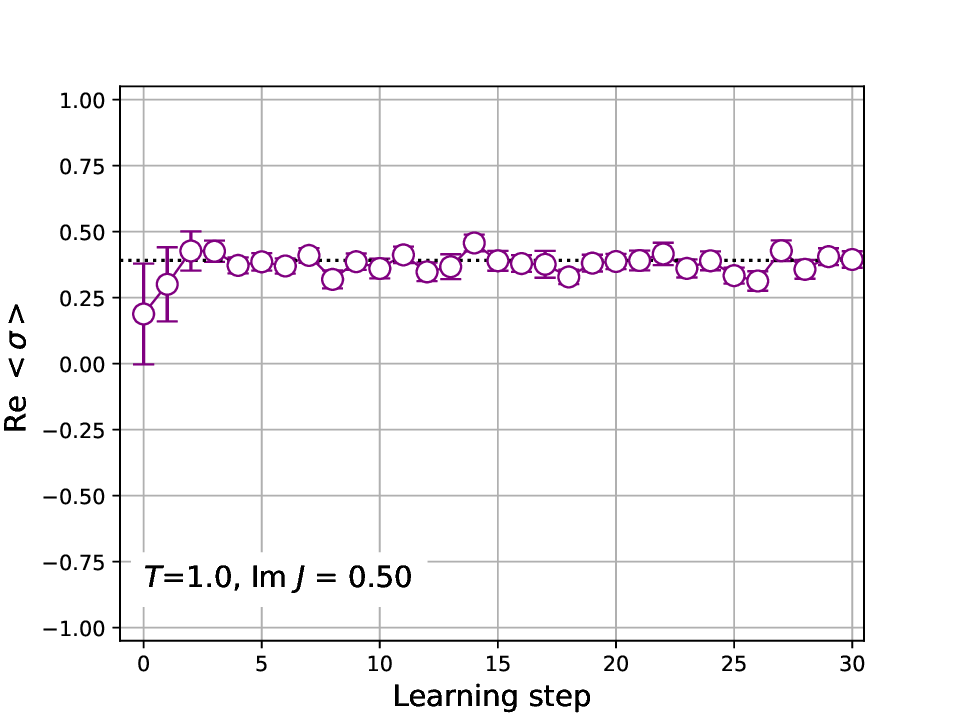}
 \includegraphics[width=0.235\textwidth]{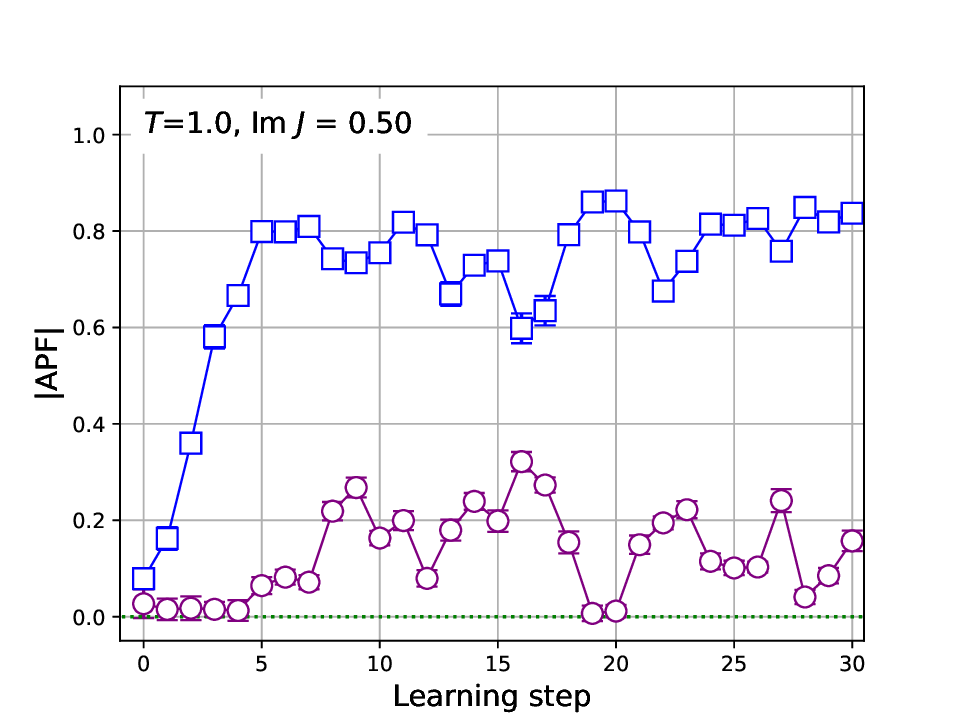}\\
 \includegraphics[width=0.235\textwidth]{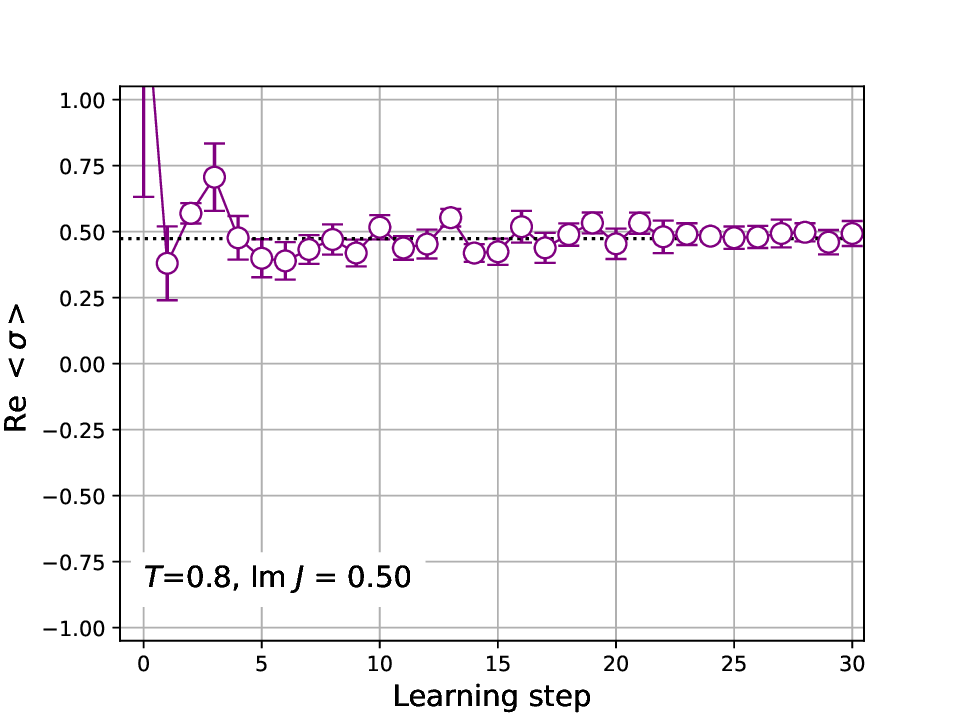}
 \includegraphics[width=0.235\textwidth]{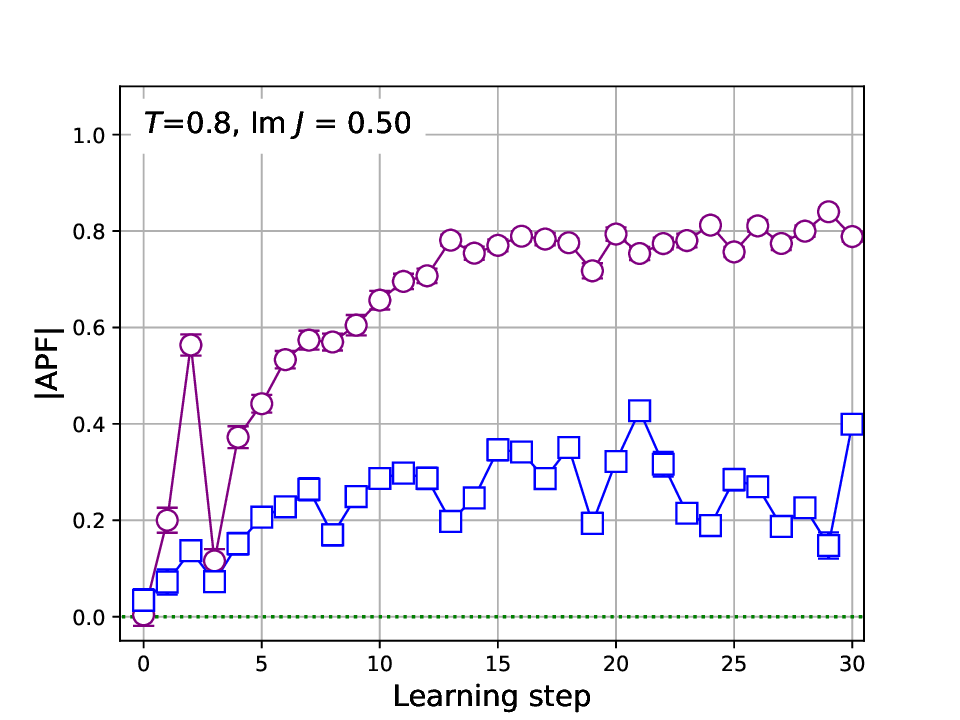}
 \caption{The magnetization and APF with $\mathrm{Im}\,J = 0.5$ at $T=0.8$, $1.0$ and $1.2$.
          The three improvements are included in the training.
          The left panel shows the real part of the magnetization and the right panel shows APF.
          The circle and square symbols in the right panel are the results of the real and imaginary parts of APF, respectively.}
\label{fig:ImJ}
\end{figure}

\begin{figure}[t]
 \centering
 \includegraphics[width=0.235\textwidth]{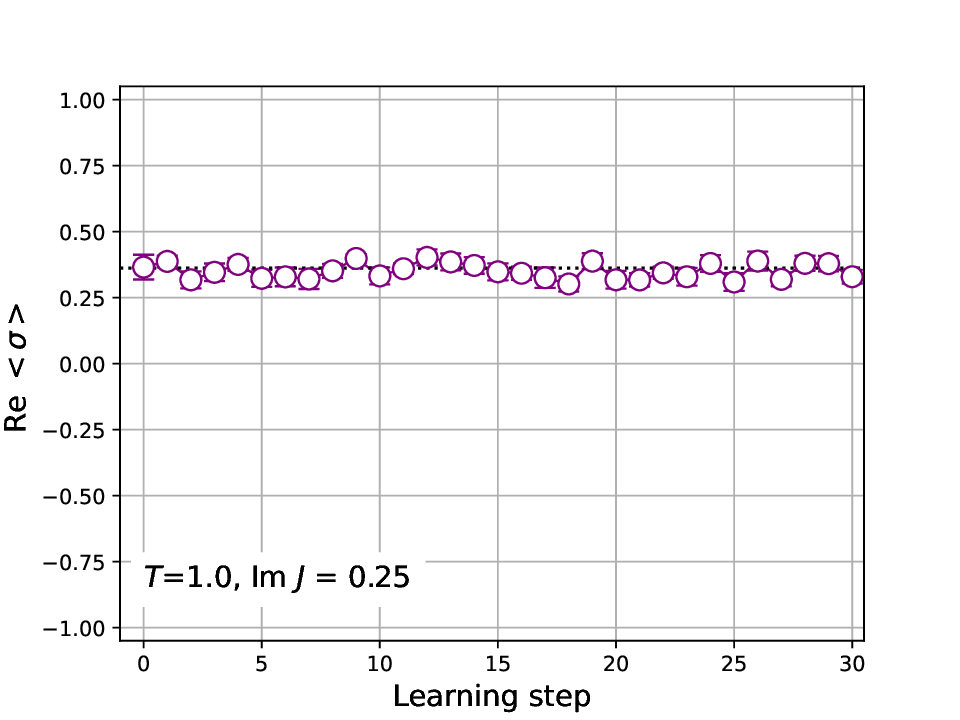}
 \includegraphics[width=0.235\textwidth]{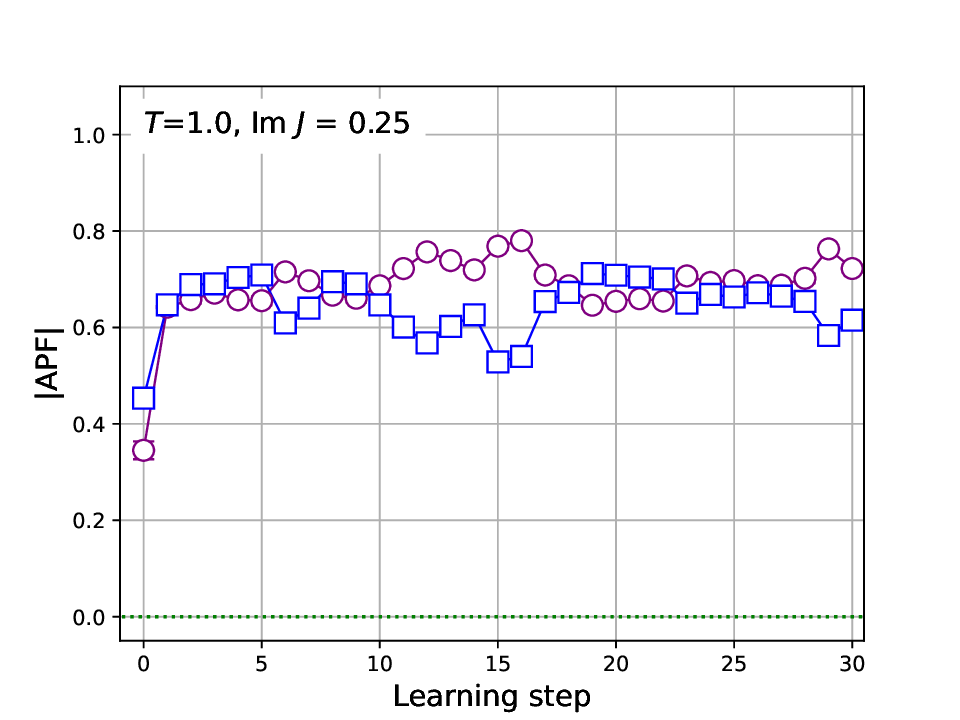}\\
 \includegraphics[width=0.235\textwidth]{Learning_step_10_050.eps}
 \includegraphics[width=0.235\textwidth]{Learning_step2_10_050.eps}\\
 \includegraphics[width=0.235\textwidth]{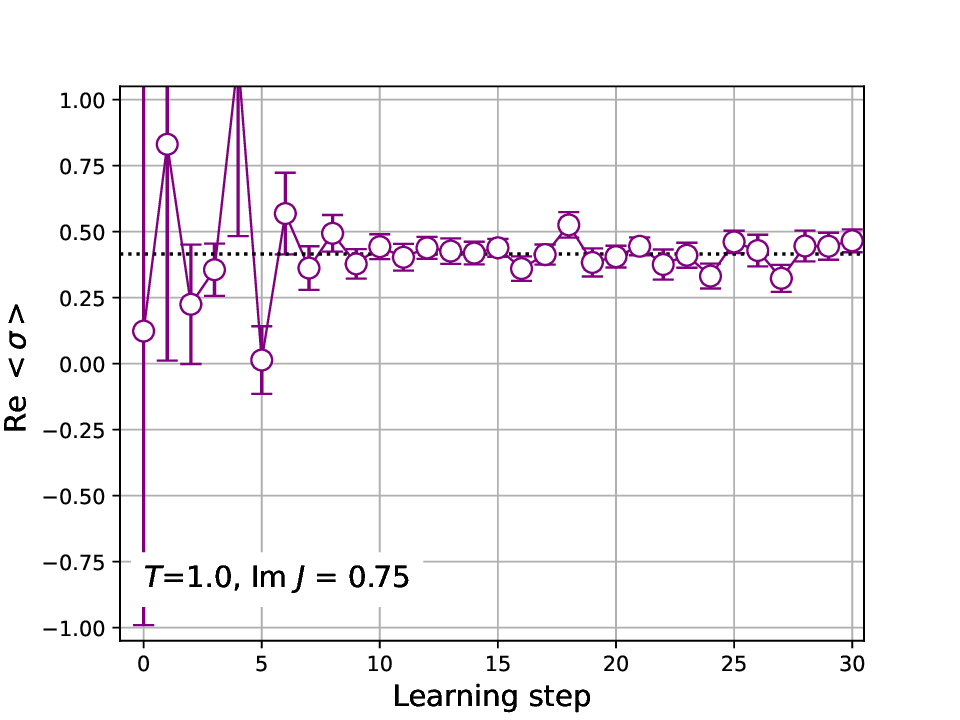}
 \includegraphics[width=0.235\textwidth]{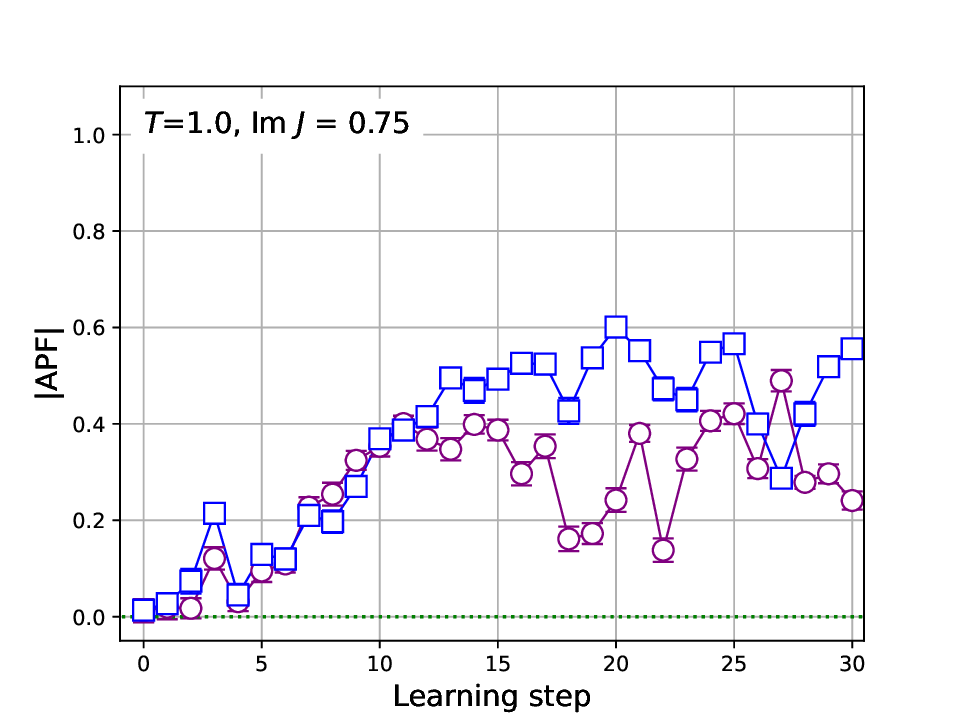}\\
 \includegraphics[width=0.235\textwidth]{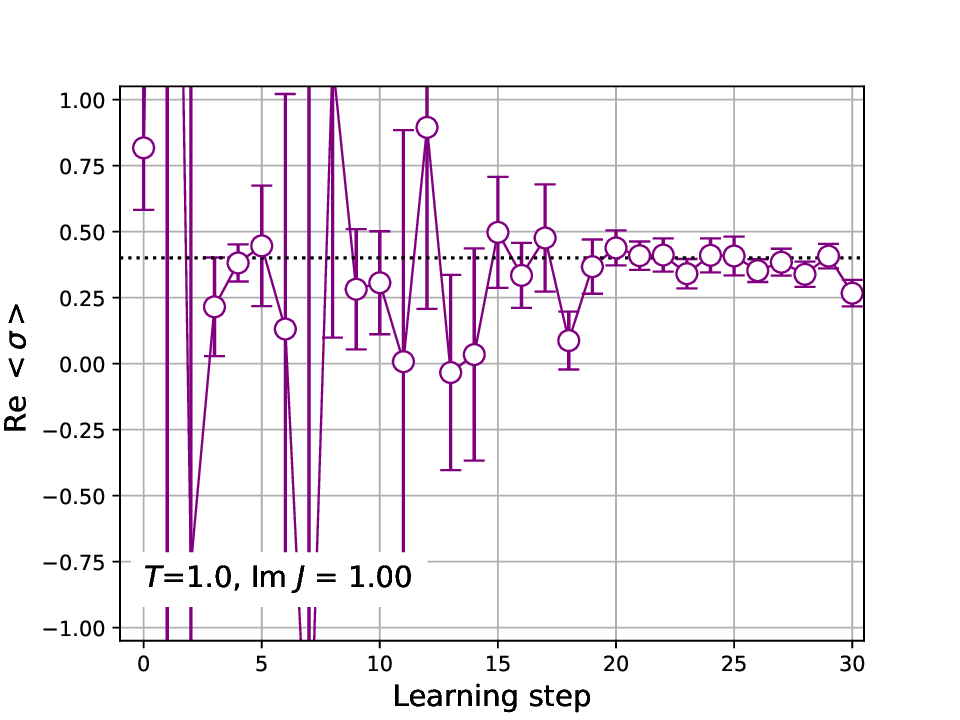}
 \includegraphics[width=0.235\textwidth]{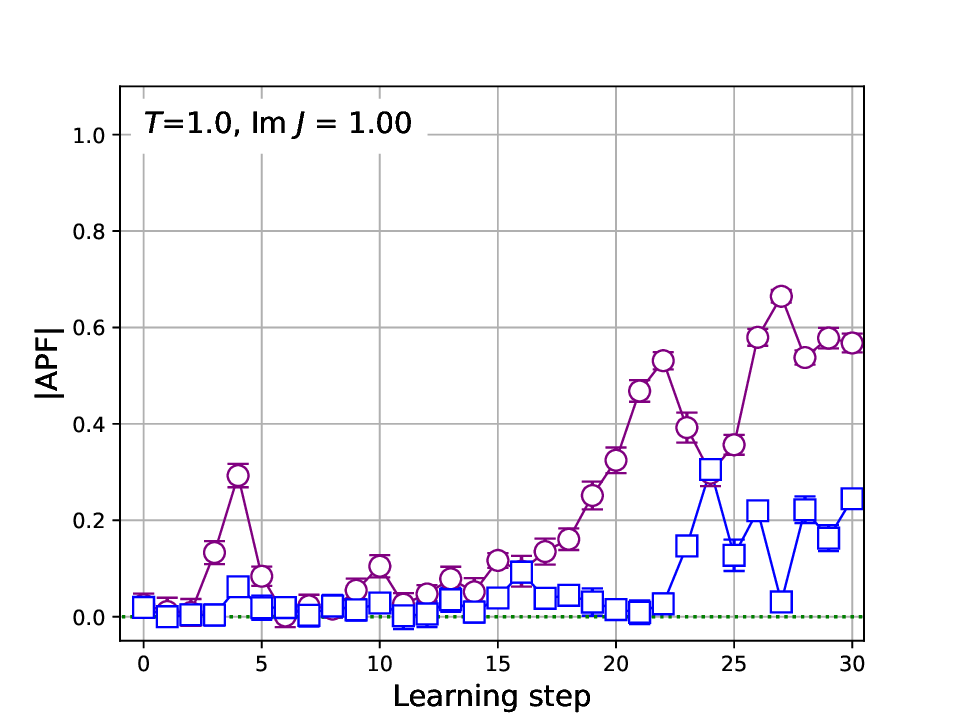}
 \caption{The magnetization and APF with $\mathrm{Im}\,J = 0.25$, $0.5$, $0.75$ and $1.0$ at $T=1.0$.
          The three improvements are included in the training.
          The left panel shows the real part of the magnetization and the right panel shows APF.
          The dotted line in the left panel denotes the analytic result. 
          The circle and square symbols in the right panel are the results of the real and imaginary parts of APF, respectively.}
\label{fig:ImJ_mod}
\end{figure}

\begin{figure}[t]
 \centering
 \includegraphics[width=0.235\textwidth]{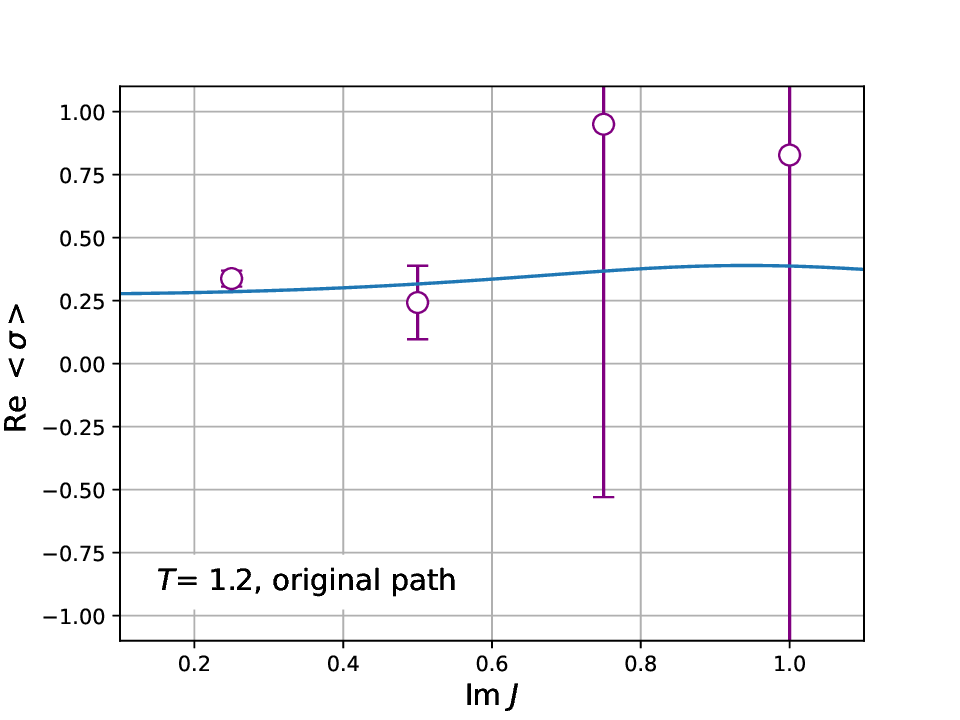}
 \includegraphics[width=0.235\textwidth]{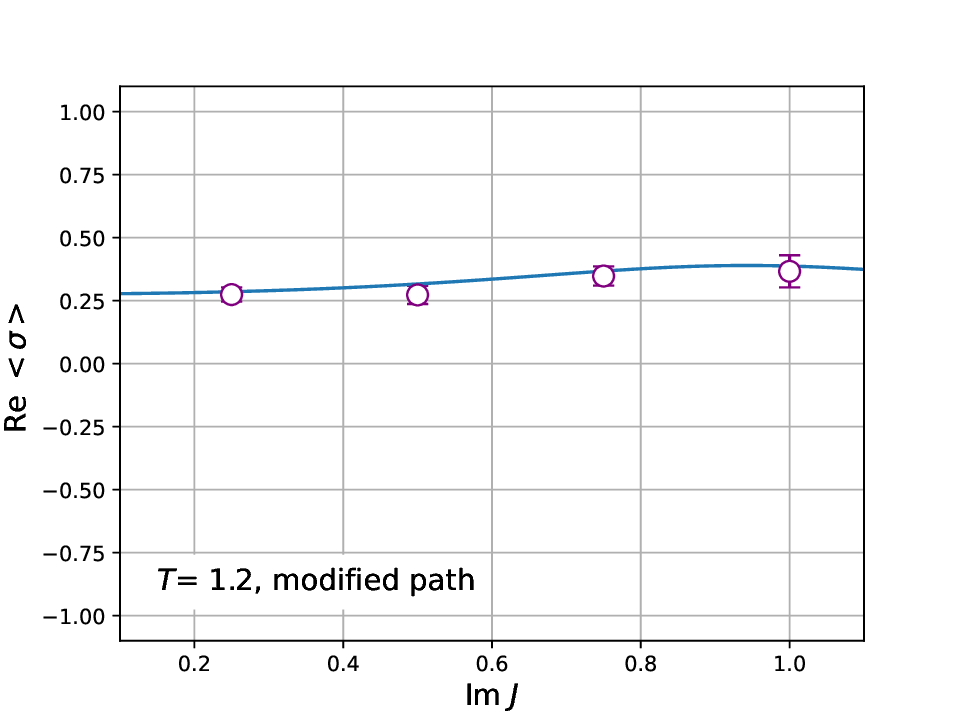}\\
 \includegraphics[width=0.235\textwidth]{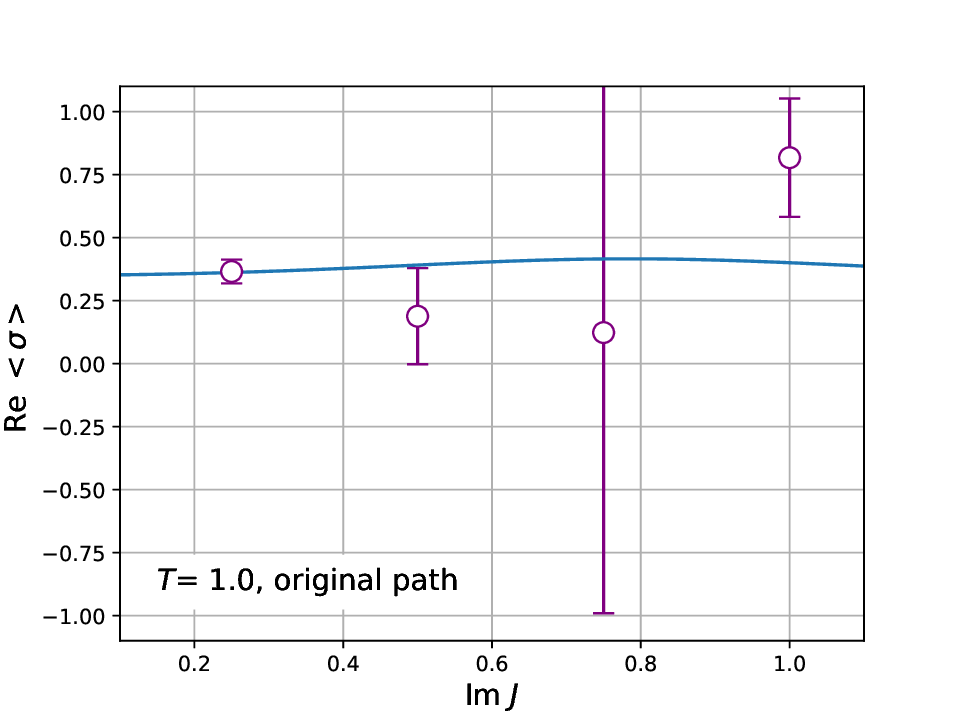}
 \includegraphics[width=0.235\textwidth]{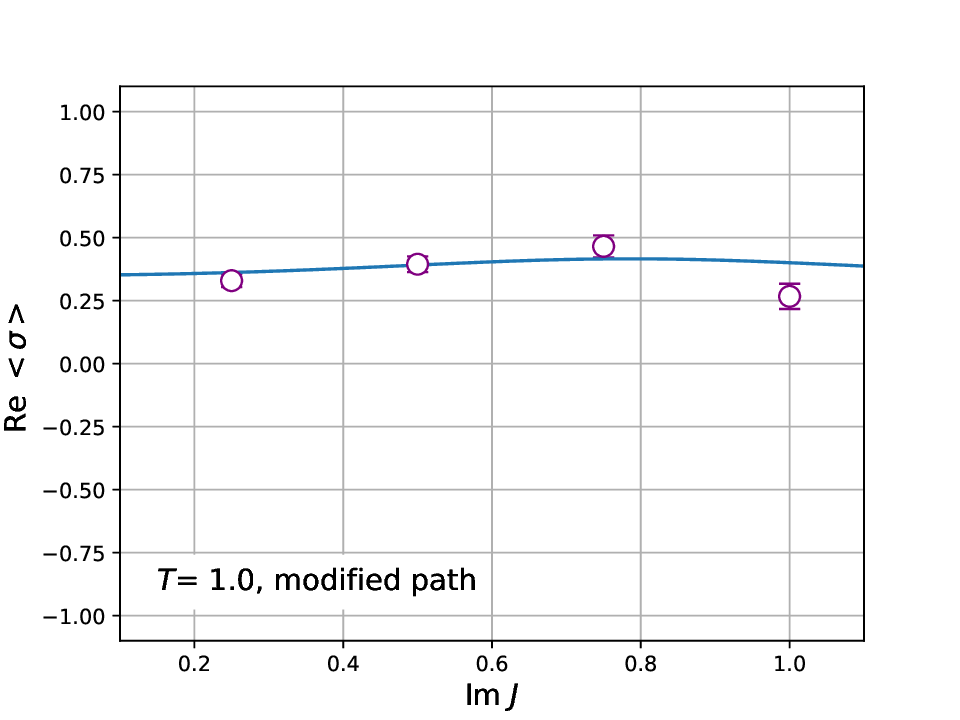}\\
 \includegraphics[width=0.235\textwidth]{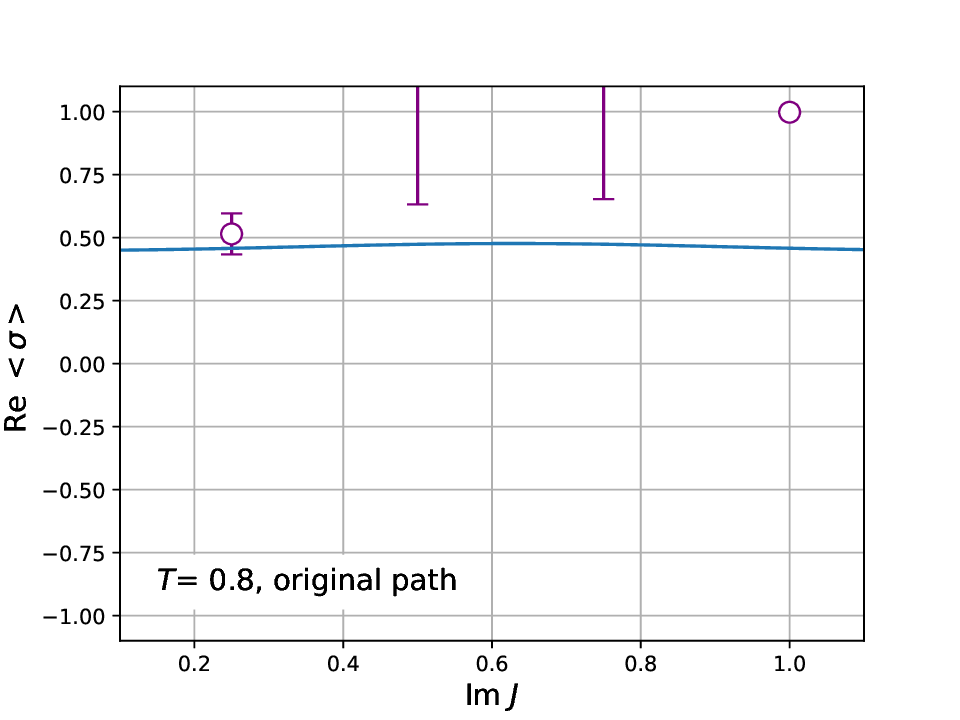}
 \includegraphics[width=0.235\textwidth]{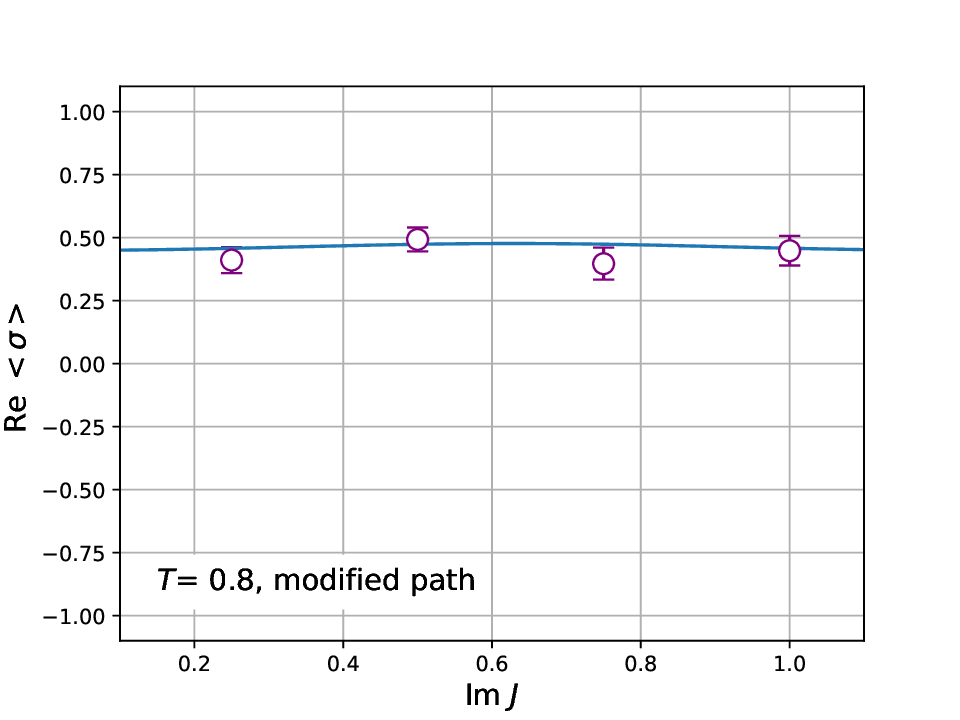}
 \caption{The magnetization with $\mathrm{Im}\,J=0.25 \sim 1.0$ at $T=0.8$, $1.0$ and $1.2$.
          The three improvements are included in the training.
          The left (right) panel is the result on the original (modified) path.}
\label{fig:ImJ-dep}
\end{figure}

Figure\,\ref{fig:ImJ-simple} shows the real and imaginary parts of APF for $T=0.8$, $1.0$ and $1.2$ with $\mathrm{Im}\,J=0.5$ for each learning step, where each learning step contains batch training and MC update.
The results in the figures are obtained without improvements explained in Sec.\,\ref{sec:extensions}.
In learning steps, the training is almost stable with large APF, but sometimes shows a sudden drop; see Appendix\,\ref{sec:histgram} for the distribution of the phase of the Boltzmann weight in the training.
This problem may be solved with a large number of replicas because the bias of sampling in HMC is relaxed.
We may also need a deeper neural network or a network based on physical knowledge of the model and/or theory to enhance the expressive power of the neural network.
We will keep them in our future work.

Figure\,\ref{fig:ImJ} shows the real and imaginary parts of APF for $T=0.8$, $1.0$, and $1.2$ with $\mathrm{Im}\,J=0.5$ for each learning step.
Comparison of Fig.\,\ref{fig:ImJ-simple} with Fig.\,\ref{fig:ImJ} suggests that training becomes more stable than that without the improvements.
Enhancement of APF is also observed in Fig.\,\ref{fig:ImJ_mod}, which shows the real part of magnetization and the real and imaginary parts of APF at fixed $T=1.0$ with $\mathrm{Im}\,J=0.25 \sim 1.0$ for each learning step.

Figure~\ref{fig:ImJ-dep} shows the magnetization on the original and modified paths.
Here, we consider $\mathrm{Im}\,J=0.25 \sim 1.0$ with $T=0.8$, $1.0$ and $1.2$.
On the original path, the statistical errors are large due to the small APF, at least in the present number of configurations.
In some regions, the error becomes very small, but the results do not reproduce the analytic result; this indicates that the HMC on the original path does not sample all relevant configurations.
On the modified path constructed by the path optimization method with some improvements, we can see that the statistical errors are well reduced.

\section{Summary}
\label{sec:summary}

In this paper, we have applied the path optimization method~\cite{Mori:2017nwj,Alexandru:2018fqp,Bursa:2018ykf} to the (Lenz-)Ising model with a complex coupling constant, which is prepared as a laboratory to investigate the sign problem in spin models with the discretized degrees of freedom.
The sum of spins is transformed into an integral using the Hubbard-Stratonovich transformation~\cite{NIPS2012_c913303f,Ostmeyer:2019dkt}, which allows us to modify the integral path on the real dynamical variable plane to that on the complex dynamical variable plane.
We found that the path optimization method works in the spin model, at least in the Ising-type model.
The average phase factor is enhanced on the modified integral path compared to that on the original integral path with improvements: the parallel tempering, the penalty term in the cost function, the mixed configurations in the training part, and the scheduler.
On the original path, the statistical error of the magnetization can be huge, or can be underestimated even with 1000 configurations indicating lack of all relevant contributions in sampling, due to the sign problem.
On the modified path by the path optimization, the expectation value of the magnetization reproduces the exact result with a well-reduced statistical error.

It should be noted that the same procedure can work also in the gauge theory case if we can rewrite the sum for spins in the path integral.
However, we should be careful with the gauge symmetry because it is hard to enhance the average phase factor without adequate treatment of the gauge symmetry.
We may need the suitable gauge fixing~\cite{Kashiwa:2020brj}, the gauge-invariant input~\cite{Namekawa:2021nzu}, or the gauge-covariant network~\cite{Tomiya:2021ywc,Namekawa:2022liz} for the path optimization method.

Since the one-dimensional Ising model does not have a phase transition, it is interesting to apply the present method to the spin model which shows a phase transition, such as the higher dimensional Ising model and also the Potts model.
While we only use machine learning to represent the integral path, we can also use it to accelerate the sampling of configurations near the phase transition point~\cite{liu2017self}.
We will report on these issues elsewhere.

\begin{acknowledgments}
This work is supported by the Japan Society for the Promotion of Science (JSPS) KAKENHI Grant Numbers
19H01898, 
21H00121, 
21K03553 and 
22H05112. 
K.K., Y.N., and H.T. would like to thank the late Prof. Akira Ohnishi without whom this project is never completed.
\end{acknowledgments}

\appendix

\section{QCD-like Potts model}
\label{sec:QCD_like_Potts}

\begin{figure*}[t]
 \centering
 \includegraphics[width=0.33\textwidth]{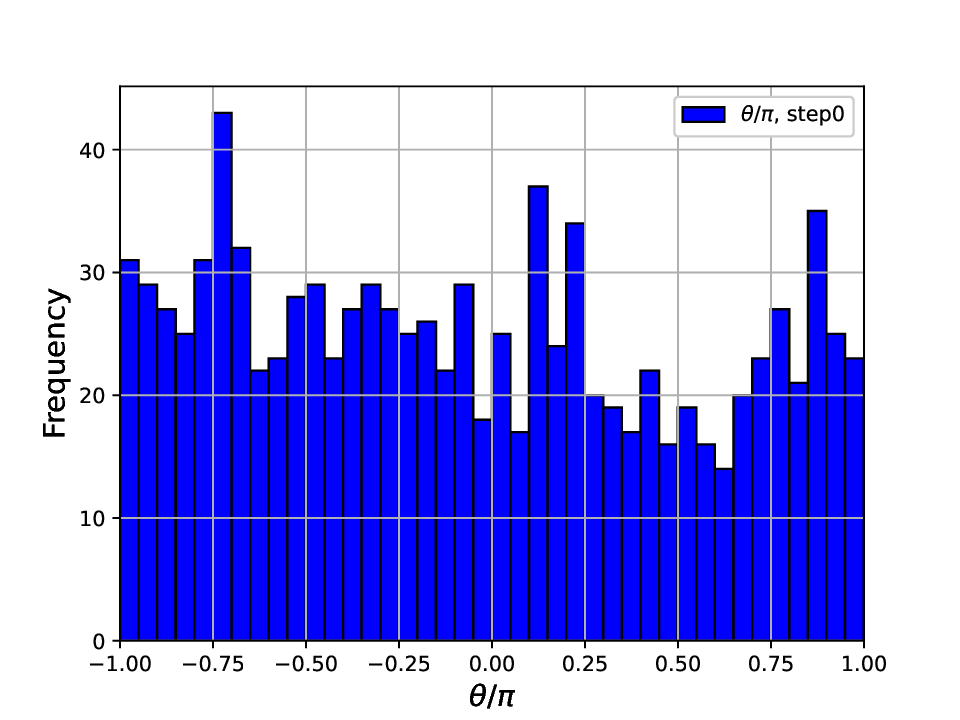}
 \hspace{-5mm}
 \includegraphics[width=0.33\textwidth]{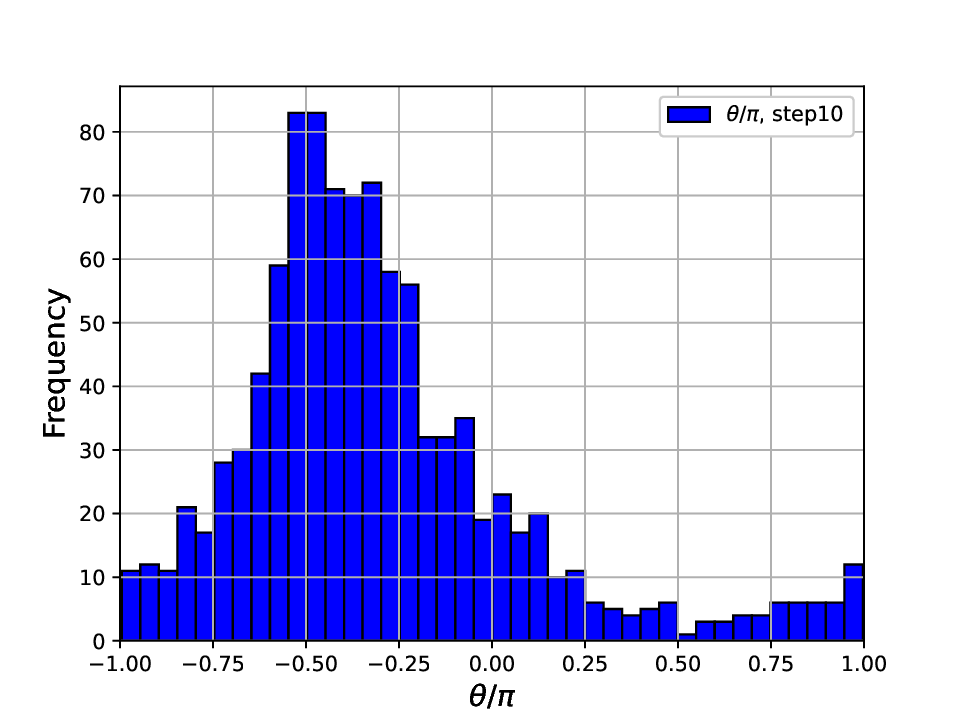}
 \hspace{-5mm}
 \includegraphics[width=0.33\textwidth]{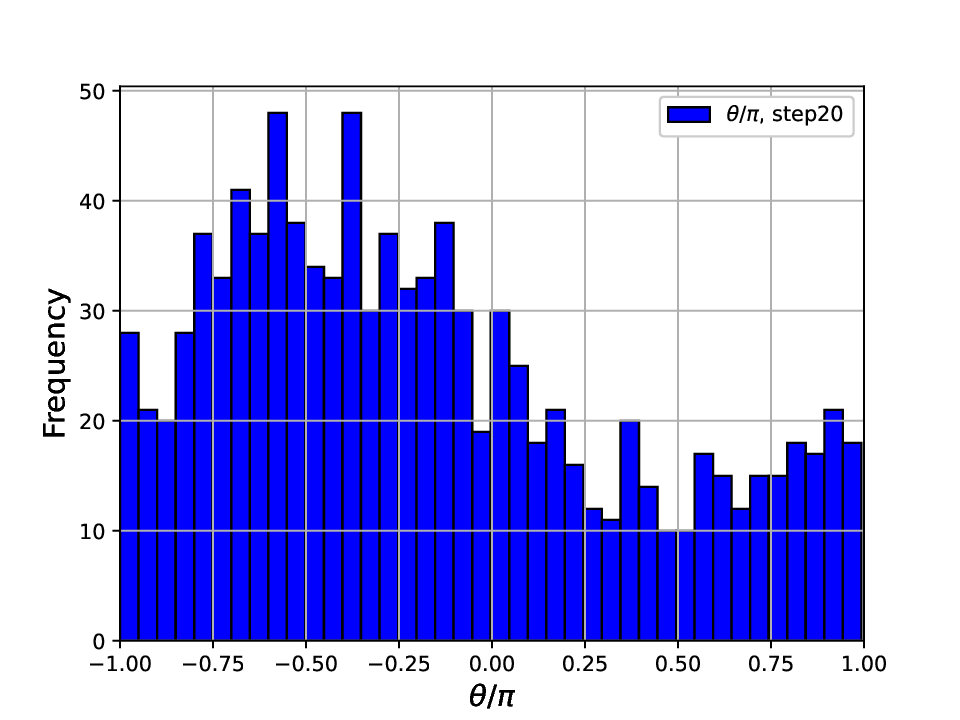}\\
 \includegraphics[width=0.33\textwidth]{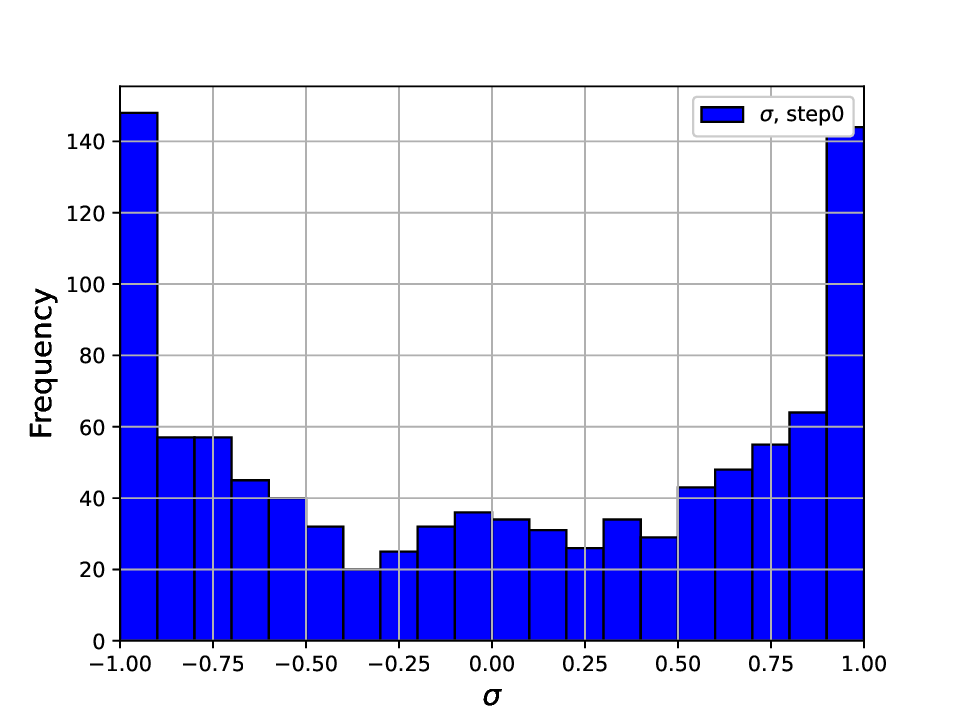}
 \hspace{-5mm}
 \includegraphics[width=0.33\textwidth]{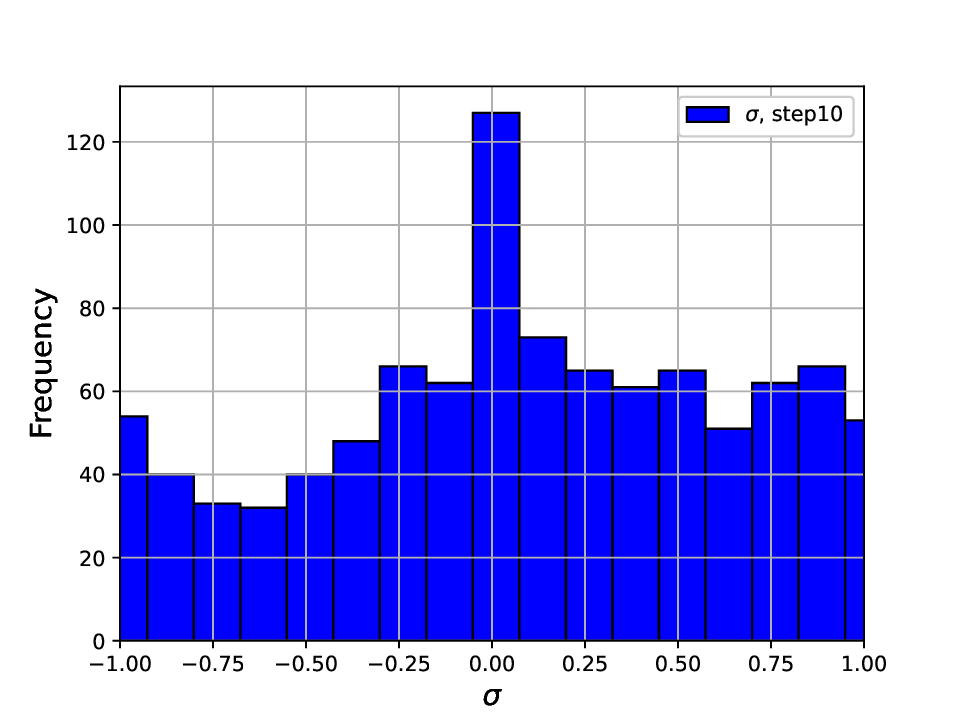}
 \hspace{-5mm}
 \includegraphics[width=0.33\textwidth]{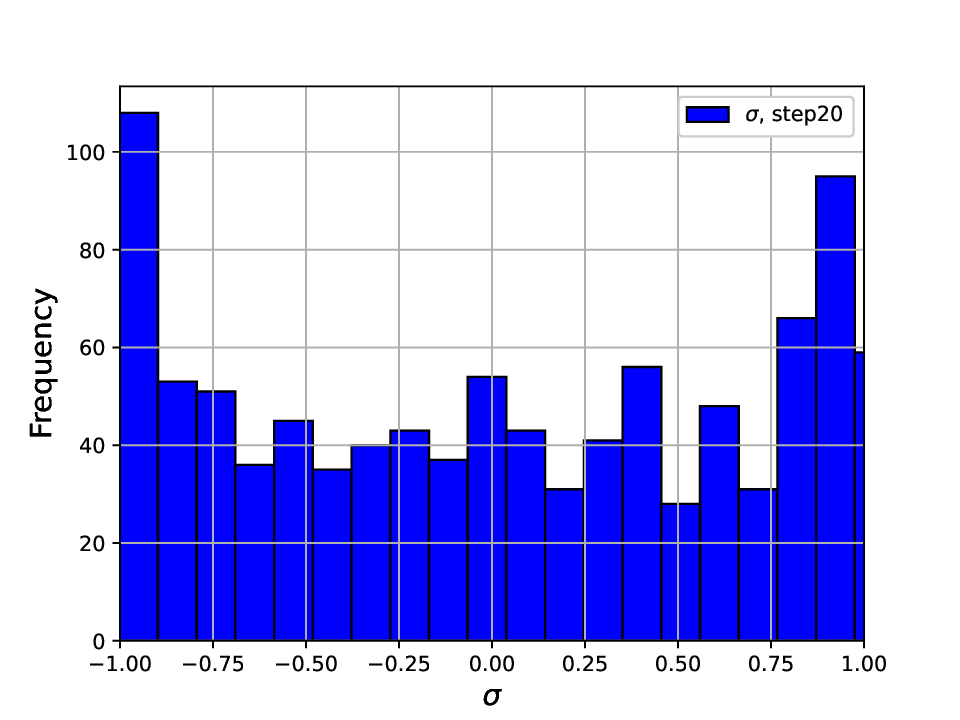}
 \caption{The top (bottom) panel shows the histogram for the phase of the Boltzmann weight (un-reweighted magnetization) with $\mathrm{Im}\,J=0.5$ at $T=1.0$ for the $0$th, $10$th and $20$th learning steps from the left to the right panel without the improvements.}
\label{fig:histgram}
\end{figure*}

We explain the possible formulation of the QCD-like Potts model as a laboratory to investigate the sign problem that appears in QCD. 
It is well known that QCD can be approximated using the Potts model if the bare quark mass is large enough.
The Hamiltonian of the QCD-like Potts model with heavy-quark contributions~\cite{Alford:2001ug,Kim:2005ck,Kashiwa:2020waa} is given by
\begin{align}
    {\cal H} &=  - \kappa  \sum_{n.n.} \delta_{k_{\bf x} k_{\bf x+i}}
          - \sum_{\bf x} \Bigl( h_- \Phi_{\bf x} + h_+ {\bar \Phi}_{\bf x} \Bigr),
\label{eq:potts}
\end{align}
where $\kappa \in \mathbb{R}$ is the coupling constant, $h_\mp \in \mathbb{R}$ means the strength of the external field, ${\bf i}$ is the unit vector for the spatial directions, and $k_{\bf x}$ are the $N_\mathrm{c}$-state Potts spin which is $\mathbb{Z}_{N_\mathrm{c}}$ quantities for each site here $N_\mathrm{c}$ denotes the number of colors;
$N_\mathrm{c}$ is set to $3$.
The last term of Eq.\,(\ref{eq:potts}) represents the heavy quark contributions.
The sum $\sum_{n.n}$ means that the spins of the nearest neighbor are summed. 
The quantity $\Phi$ (${\bar \Phi}$) is called the Polyakov loop (its conjugate) in QCD, and is defined as
\begin{align}
    \Phi_{\bf x} = \exp \Bigl( {\frac{2\pi i k_{\bf x}}{N_\mathrm{c}}} \Bigr), ~~~~
    {\bar \Phi}_{\bf x} = \exp \Bigl( -{\frac{2\pi i k_{\bf x}}{N_\mathrm{c}}} \Bigr).
\end{align}
The functional form (\ref{eq:potts}) was first shown in Ref.\,\cite{Alford:2001ug} and was further discussed in Ref.\,\cite{Kim:2005ck}.
Detailed discussions of the form of $h_\mp$ are shown in Ref.\,\cite{Kashiwa:2020waa}.
The strength of the external field can be expressed as
\begin{align}
    h\mp = e^{-\beta (M \mp \mu)},
\end{align}
where $\beta$ is the inverse temperature $\beta = 1/T$, $M$ denotes the bare quark mass and $\mu$ means the quark chemical potential; the bare quark mass must be large enough compared with some other energy scales, in principle.
If the nearest-neighbor spins take the same value, the first term decreases the energy.
At zero temperature, the spin-aligned state is favored if $\kappa > 0$.
To investigate the detailed structure of the Ising model and Potts model, the Metropolis method~\cite{metropolis1953equation} is widely used if there is no sign problem.

Equation (\ref{eq:potts}) is the simplest form of the QCD-like Potts model, but it is not suitable for the path optimization.
We thus modify the first term as
\begin{align}
    \kappa \sum_{n.n.} \delta_{k_{\bf x} k_{\bf x+i}}
    \to \frac{\kappa}{2} \sum_{n.n.} \Bigl( \Phi_{\bf x} {\bar \Phi}_{\bf x+i} + {\bar \Phi}_{\bf x} \Phi_{\bf x+i} \Bigr).
\end{align}
This replacement does not significantly change the properties of the Potts model.
The energy is decreased if the nearest-neighbor spins take the same value even with the right-hand side term.
This form is similar to a part of the Polyakov loop potential and therefore may be suitable for the QCD-like model; see Refs.\,\cite{Ratti:2005jh,Fukushima:2017csk}.
The total energy becomes
\begin{align}
    {\cal H} &=  - \frac{\kappa}{2} \sum_{n.n.} \Bigl( \Phi_{\bf x} {\bar \Phi}_{\bf x+i} + {\bar \Phi}_{\bf x} \Phi_{\bf x+i} \Bigr)
          - \sum_{\bf x} \Bigl( h_- \Phi_{\bf x} + h_+ {\bar \Phi}_{\bf x} \Bigr)
    \nonumber\\
    & = - \frac{1}{2} s^\top A s - h s,
\label{eq:potts2}
\end{align}
where $s$ is a complex vector with $2N$-components consisting of $\Phi_{\bf x}$ and ${\bar \Phi}_{\bf x}$, $A$ is the $2N \times 2N$ symmetric matrix, and $h=(h_-,\cdots,h_-,h_+,\cdots,h_+)$ where $N$ is the number of sites.
We impose the periodic boundary condition for this model.

To keep the first term in Eq.\,(\ref{eq:potts2}) real, we sum up a possible combination of $\Phi$ and ${\bar \Phi}$.
The partition function is then given by
\begin{align}
    {\cal Z} = \sum_{\{k\}} e^{-{\cal H}},
\end{align}
where the sum takes over all possible states of the Potts spins and $\beta$ is absorbed into ${\cal H}$.
Since the expression (\ref{eq:potts2}) is similar to that of the Ising model , we can use the same formulation.
Therefore, we can use the hybrid Monte Carlo method for the QCD-like Potts model if the effective Hamiltonian is real.
The Hamiltonian becomes complex at finite $\mu$, which causes the sign problem.
It should be noted that the expectation value of the energy must be positive.

Since degrees of freedom in the present model can be expressed by continuous dynamical variables, the path optimization method can be applied to the QCD-like Potts model.

\section{Histogram of Boltzmann weight}
\label{sec:histgram}

We show the histogram of the phase of the Boltzmann weight in some learning steps for the model without using the penalty term, mixed configurations, and the scheduler.
Figure\,\ref{fig:histgram} shows the phase of the Boltzmann weight with $\mathrm{Im}\,J=0.5$ at $T=1.0$ for the $0$th, $10$th, and $20$th learning steps.
After the $10$th learning step, the APF suddenly drops, as shown in Fig.\,\ref{fig:ImJ-simple}.
The phase distributions at the $0$th and $20$th learning steps are not very localized, unlike the $10$th learning step, which can be the source of the small APF.
The un-reweighted magnetization also shows the two peak structures when the training result is not good.
There may be flat directions which obstruct the training.

\bibliography{ref.bib}

\end{document}